\documentclass[twocolumn,twoside]{article}
\renewcommand{\baselinestretch}{1.05}
\makeatletter
\renewcommand{\theequation}{%
	\thesection.\arabic{equation}}
	\@addtoreset{equation}{section}
\makeatother

\input epsf.tex

\begin{document}
\title{Interplay between elastic fields due to gravity and a partial dislocation
for a hard-sphere crystal coherently grown under gravity: driving force for defect
disappearance}
\author{Atsushi Mori\dag\thanks{Corresponding author. E-mail: mori@opt.tokushima-u.ac.jp\vspace{6pt}},
Yoshihisa Suzuki\ddag}
\date{
\begin{center}
\begin{small}
\dag{}Department of Advanced Materials,
Institute of Technology and Science,
The University of Tokushima,
2-1 Minamijosanjima, Tokushima 770-8506, Japan \\
\ddag{}Department of Life System,
Institute of Technology of Science,
The University of Tokushima,
2-1 Minamijosanjima, Tokushima 770-8506, Japan \\[1ex]
(\textit{recieved\hspace{3cm}; in final form\hspace{3cm}})
\end{small}
\end{center}}
\maketitle
\begin{small}
We previously observed that an intrinsic staking fault shrunk through a glide
of a Shockley partial dislocation terminating its lower end in a hard-sphere
crystal under gravity coherently grown in $\langle$001$\rangle$ by Monte Carlo simulations
[Mori~{\it et~al.}, Molec.~Phys. {\bf 105}, 1377 (2007)];
it was an answer to a one-decade long standing question why the stacking disorder
in colloidal crystals reduced under gravity
[Zhu~{\it et~al.}, Nature {\bf 387}, 883 (1997)].
Here, we present an elastic energy calculation; in addition to the
self-energy
of the partial dislocation [Mori~{\it et~al.}, Prog.~Theor.~Phys.~Suppl.
{\bf 178}, 33 (2009)] we calculate the cross-coupling term between elastic
field due to gravity and that due to a Shockley partial dislocation.
The cross term is a increasing function of the linear dimension $R$
over which the elastic field expands, showing that a driving force arises
for the partial dislocation moving toward
the upper boundary of a grain. 
\end{small}
\section{Introduction \label{sec:intro}}
Zhu~{\it et~al.} \cite{Zhu1997} in 1997 presented results of colloidal
crystallization, which implied that the stacking disorder in the colloidal
crystals reduces due to gravity.
They made such conclusion by comparison between results of
a Space Shuttle experiment and those on the ground that whereas under micro
gravity the colloidal crystals are of random hexagonal-close pack (rhcp) structure,
under normal gravity they exhibit a mixture of rhcp and the face-centered
cubic (fcc) crystal.
In a previous experiment the structure under normal gravity was fcc \cite{Pusey1989}.
This trend was supported by Kegel and Dhont \cite{Kegel2000}.
Their result was, however, slightly different; the structure was faulted
twinned fcc under normal gravity.
In their experiments the hard sphere (HS) nature was realized
by using poly(methyl methacrylate) (PMMA) microspheres suspended in
a hydrocarbon medium \cite{Antl1986}.
The difference in the particle size and dispersion medium may results
in the variety of the final states under the normal gravity.
We point out, however, that the variety is nature.
It is natural to understand that the non-uniqueness of the final states under
gravity results from metastable equilibria.
Processes of disappearing stacking disorder must be those between certain
metastable states.

We performed Monte Carlo (MC) simulations of HSs at various gravitational
conditions to elucidate the mechanism of the defect disappearance due to gravity.
The stacking sequence of the hexagonal planes does not affect the density.
So, the mechanism has been a hot question since Zhu~{\it et~al.} \cite{Zhu1997}.
It was found in this simulation that a defective crystal was formed above
a less-defective crystalline region at the bottom of the system \cite{Mori2006JCP}.
The lateral cross sections were square.
Therefore, in case that the lateral system size was small{\bf,}
(001) planes were stacked along vertical direction.
On the other hand, in case that the lateral system size was not so small,
until the strength of gravity was not so large that the packing of particles
at the bottom was not dominated by the smallness of the system (111) planes
were stacked.
One can understand those phenomena as results from competition between
contributions of the interfacial free energy against bottom wall and stress
from the lateral boundary.
We wish to point out that stress equivalent in those simulations can
be given by use of a patterned substrate; the use of a patterned substrate
was proposed by van~Blaaderen~{\it et~al.} \cite{Blaaderen1997}.
This method is called a colloidal epitaxy.
We have taken a detailed look into processes in the (001) growth,
in which a portion of the defective region is transformed
into the less defective state \cite{Mori2007MP}.
In the (001) growth \{111\} planes and, thus, stacking faults run along
a oblique direction, not horizontal.

From illustrations ({\it e.g.} Fig.~1)
we have an intuition that the strain due to a Shockley partial dislocation
at the lower end of an stacking fault cooperates with gravitational effect
toward defect disappearance.
To give support to this intuition we have performed elastic theoretical
calculations.
A part of them has already been presented \cite{Mori2009};
the self-energy of the partial dislocation as well as the buoyancy
due to particle deficiency of the core structure yields a driving force
for the partial dislocation moving upward.
Interfacial energy due to stacking disorder has been added as well as
the dislocation core energy.
The former is a form of the diving force for shrinking of a staking fault
running upward.
Section~\ref{sec:self-energy} repeats some details of the elastic theoretical
calculation presented already \cite{Mori2009}.

Strain energies due to a patterned substrate were evaluated for wetting
by a crystalline layer \cite{Heni2000,Heni2001}.
Also strain energy was estimated for misfit dislocation in colloidal
epitaxy \cite{Schall2004}.
The critical thickness $h_c$ for misfit dislocation was estimated and
for colloidal crystals thicker than $h_c$ a position where the stacking
fault starts upward was given.
Comparison between strain energies due to the substrate and due to lateral
boundary would answer whether in phenomena in our MC simulation under small
system size corresponds to those in colloidal epitaxy.
Strain in our MC simulation is geometrically equivalent to those which
appear due to patterned substrate \cite{Mori2006STAM}.
The interplay between gravitational and substrate effects
is left for a future research.
We will concentrate in this paper on the stability of an intrinsic
stacking fault terminated by a Shockley partial dislocation.

\section{System \label{sec:system}}
\begin{figure}[t]
\centerline{
\epsfxsize=0.5\textwidth
\epsfbox{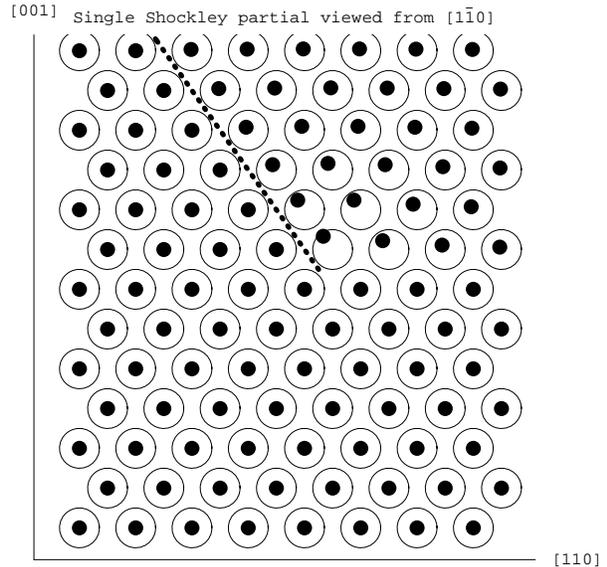}}
\caption{\label{fig:single}
Illustration (not a snapshot of a simulation)
of an intrinsic stacking fault which is disappearing.
Particles in the distorted crystal are shown by the dots and the regular
lattice positions by open circles.
A dotted line indicates the stacking fault.
In this illustration particles outside the portion right of this line are fixed
at the regular positions.
The intrinsic stacking fault is terminated by a Shockley partial dislocation
at the center of this illustration.}
\end{figure}
An intrinsic stacking fault disappearing
is schematically shown in an illustration (Fig.~1).
It is seen that a Shockley partial dislocation of the Burgers vector
$\mbox{\boldmath $b$}^I = (1/6)[\bar{2}11]$
($\equiv -\mbox{\boldmath $a$}_1/3 + \mbox{\boldmath $a$}_2/6
+ \mbox{\boldmath $a$}_3/6$ with $\mbox{\boldmath $a$}_1$,
$\mbox{\boldmath $a$}_2$, and $\mbox{\boldmath $a$}_3$ being the lattice vectors)
terminates the stacking fault.
Translation of (111) plane by $\mbox{\boldmath $b$}^I$
shifts the positions of particles in the (111) plane from A, B, and C to B, C,
and A, respectively (see, Fig.~2.)
Here, A, B, and C refer to three possible positions in projection onto the (111)
plane.
Pictures similar to Fig.~1
obtained by confocal microscopy have been presented
by Schall~{\it et~al.} \cite{Schall2004} and Alsayed~{\it et~al.} \cite{Alsayed2005}.
Phenomena observed by Schall~{\it et~al.} are, however, different from what
we focus in this paper.
They have focused on misfit dislocations and have observed increasing dislocations.
On the other hand, Alsayed~{\it et~al.} observed premelting near varieties of defects.
Though the purposes are different, the Shockley partial dislocations evidently exist
in colloidal crystals.
The stacking fault runs upward and the linear dimension $R$ over which the elastic field
expands is regarded as the distance from the Shockley
partial dislocation to the upper boundary of the crystalline grain.
We will not consider the interaction between the bottom substrate and the Shockley
partial dislocation.

\begin{figure}[t]
\centerline{
\epsfxsize=0.5\textwidth
\epsfbox{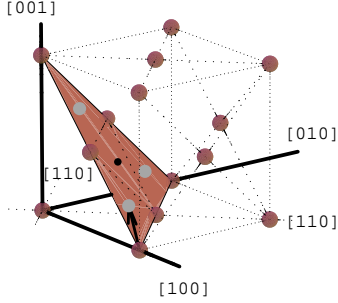}}
\caption{\label{fig:Shockley}
The Burgers vector $\mbox{\boldmath $b$}^I = (1/6)[\bar{2}11]$ (arrow)
and (111) plane (painted).
The arrow connects a lattice position at, say, A and an adjacent one at, say, B.}
\end{figure}

We note here that the density decreases with the altitude $z$, according to
the mechanical equilibrium equation
\begin{equation}
\label{eq:mechanical}
\frac{\partial P}{\partial z} = -mg\rho(z),
\end{equation}
where $P$ is the pressure, $m$ the mass of a particle, $g$ the acceleration
due to gravity, and $\rho(z)$ the particle number density at $z$ in coarse scale.
The other point characteristic of these simulations is that the coherent
growth \cite{Mori2006STAM} occurred as already mentioned.
We find that the lattice constant $c$ is linearly increases with
the altitude $z$ whereas the lattice constants $a=b$ are constant.
We can write
\begin{equation}
\label{eq:c}
c(z) = c(0)[1+\xi z],
\end{equation}
with $\xi$ being the coefficient of order $10^{-3}$ \cite{Mori2006STAM}.
In section \ref{sec:crossterm} we integrate
$f_\alpha^{(g)} = \partial \sigma_{\alpha \beta} /\partial x_\beta$
to obtain the stress $\sigma_{\alpha \beta}^{(g)}$ due to gravity
with use of Eqs. (\ref{eq:mechanical}) and (\ref{eq:c}),
where $f_{\alpha}^{(g)} = \partial P /\partial x_\alpha$.
Here, the Greek alphabets, $\alpha$, $\beta$,
have represented the Cartesian coordinate $x$, $y$, and $z$.
We will follow this convention hereafter.

\section{Self energy, buoyancy, and stacking fault energy \label{sec:self-energy}}
We shall reproduce the calculation of the sum of strain, stacking fault, and gravitational
energies for the configurations illustrated in Fig.~1
per unit length ($\sigma$) depth perpendicular to the paper,
where $\sigma$ is the HS diameter.
In Ref.~\cite{Mori2009} the elastic strain energy for an intrinsic stacking fault
was calculated by using a formula
$W(R) = (\mu b^2 / 4 \pi) [\cos^2 \theta + \sin^2 \theta / (1-\nu)] \ln (\alpha R/ b)$,
where $\mu$, $\nu$, and $\theta$ are the shear modulus, the Poisson ratio,
and the angle between the Burgers vector $\mbox{\boldmath $b$}$ and the
sense direction of the dislocation line, respectively, and $R$
is the linear dimension over which the elastic field expands and 
$r_0 \equiv b/\alpha$ defines the radius of dislocation core with
$\alpha \sim 0.1$ \cite{Hirth} (Do not confuse the coefficient $\alpha$ with
the subscript $\alpha$).
For the intrinsic stacking fault we put $\mbox{\boldmath $b$} = \mbox{\boldmath $b$}^I$
($|\mbox{\boldmath $b$}^I| = a/\sqrt{6}$ with $a$ being the fcc lattice constant)
and $\theta = \pi/6$ for $W(R)$ to obtain
\begin{equation}
\label{eq:Uel}
U_{el} = \frac{\mu a^2}{96\pi} \left(
3+\frac{1}{1-\nu} \right)
\ln \left( \frac{\sqrt{6} \alpha R}{a}
\right).
\end{equation}

We have $U_{sf} = \gamma_i l_{sf}$ for the stacking fault energy of the intrinsic
stacking fault, where $\gamma_i$ is the interfacial energy of the intrinsic stacking
fault and $l_{sf}$ is the effective length of the stacking fault.
$l_{sf}$ is the same order of $R$; a geometrical factor should be multiplied to $R$.
Pronk and Frenkel \cite{Pronk1999} calculated the stacking fault energy $\gamma_{sf}$
for the interfacial free energy between fcc and hcp crystals.
$\gamma_i$ is regarded as the same order of $\gamma_{sf}$.
A factor should be multiplied to $\gamma_{sf}$.
Introducing a factor $\zeta$ we can write the stacking fault energy as
$U_{sf} = \zeta \gamma_{sf} R$.

Let us estimate the gravitational energy.
Particle deficiency due to one Shockley partial dislocation corresponds
to one-third lattice line in Fig.~1.
A simple estimation is $U_g = (1/2\sqrt{2}) mg\rho a^2R/3$,
where $ 1/2\sqrt{2}$ comes from the number of particles per unit length
of the lattice line [110]~\footnote{This factor has been lacking in
Ref.~\cite{Mori2009}. Also, $g$ has been missing after $m$ there.}.

The sum of $U_{sf}$ and $U_g$ yields a linear term
\begin{equation}
\label{eq:linear}
U_{sf}+U_g = (\zeta \gamma_{sf} +mg \rho a^2/6\sqrt{2}) R.
\end{equation}
Because $\gamma_{sf} \sigma^2/k_BT$ calculated \cite{Pronk1999} is of order
$10^{-4}$, $U_{sf}$
is negligible small as compared to $U_g$
(in MC simulations \cite{Mori2006JCP,Mori2007MP,Mori2009,Mori2006STAM} we focus
on the phenomena at $mg\sigma/k_BT \sim 1$), where $k_BT$ is the temperature
multiplied by Boltzmann's constant.
Moreover, $\mu \sigma^3/k_BT$ = 50 $\sim$ 100 \cite{Frenkel1987,Runge1987,Laird1992};
hence, $U_{el}$ dominates.

Energy due to the dislocation core should be added.
It depends on the core structure.
In particular, in the present case the entropy term contributes
through the change in the free volume or vibrational mode due to
the dislocation core.
Hence, the accurate evaluation of core energies should be made with simulations.
Let us borrow the argument for solid states;
the core energy is proportional to $\mu b^2$.
In most cases the metal core energy $U_{core}$ is about ten percents
of $U_{el}$.
Let us write $U_{core} = \beta \mu b^2$ with $\beta$ being a certain constant
of order less than unity (Do not confuse this coefficient with the subscript
$\beta$).
Putting $\mbox{\boldmath $b$} = \mbox{\boldmath $b$}^I$ for the intrinsic
stacking fault we have
\begin{equation}
\label{eq:core}
U_{core} = \beta \mu a^2/6.
\end{equation}

The elastic constants for the HS crystal were calculated by Frenkel
and Ladd \cite{Frenkel1987} by molecular dynamic simulations,
by Runge and Chester \cite{Runge1987} by a MC simulation, and
by Laird \cite{Laird1992} by a density functional theory.
The shear modulus $\mu$ for the HS crystal ranges between 50 and 100
as already mentioned (in the unit of $k_BT/ \sigma^3$), which depends
on the particle number density $\rho$.
This range of $\mu$ corresponds to $\rho \sigma^3 \cong 1.06 - 1.13$
($a/\sigma$ $\cong$ 1.55 -- 1.52) around the disappearing stacking
disorder in the MC simulations
\cite{Mori2006JCP,Mori2007MP,Mori2009,Mori2006STAM}.
Stacking fault energy was calculated by MC simulation by Pronk and
Frenkel \cite{Pronk1999}.
They calculated interfacial free energy between fcc and hcp at
$\rho \sigma^3 = 1.10$ to be (26$\pm$6) $\times 10^{-5} k_BT/\sigma^2$,
as mentioned already.
We neglect difference among fcc-hcp, intrinsic (such as ...CAB-ABC...),
and others as done \cite{Pronk1999}.
We take a distance between the Shockley partial dislocation to the fluid region
in the simulation or a linear dimension of the grain in experiment
for the effective length of the intrinsic stacking fault $l_{sf}$ as well as $R$.
The ^^ ^^ total" energy of the intrinsic stacking fault is
\begin{eqnarray}
\label{eq:Usingle}
\nonumber
&&
U = \left[ \frac{1}{96\pi} \left( 3+\frac{1}{1-\nu} \right)
\ln \left( \frac{\sqrt{6} \alpha R}{a} \right)
+ \frac{\beta}{6} \right] \mu a^2 \\
&&
+ \left( \zeta \gamma_{sf}
+ \frac{mg\rho a^2}{6\sqrt{2}} \right) R.
\end{eqnarray}
We note here that though the materials parameters for the HS system have
been used, the formulas, Eqs.~(\ref{eq:Uel}-\ref{eq:Usingle}), Eq.~(\ref{eq:Ute})
and those derived therefrom, are valid for any materials as long as the lattice
constant varies linearly in vertical direction.

It is instructive to write the true total elastic energy in terms of the elastic
fields excluding the one which is forcing the coherent growth.
\begin{eqnarray}
\nonumber
&&
U_{t.e.} = \frac{1}{2} \int
(\sigma_{\alpha\beta}^{(edge)} + \sigma_{\alpha\beta}^{(screw)}+ \sigma_{\alpha\beta}^{(g)}) \\
\nonumber
&&
\times
(\epsilon_{\alpha\beta}^{(edge)} + \epsilon_{\alpha\beta}^{(screw)} + \epsilon_{\alpha\beta}^{(g)}) dV \\
\nonumber
&&
= \frac{1}{2} \int \sigma_{\alpha\beta}^{(edge)} \epsilon_{\alpha\beta}^{(edge)} dV \\
\nonumber
&&
+ \frac{1}{2} \int \sigma_{\alpha\beta}^{(screw)} \epsilon_{\alpha\beta}^{(screw)} dV \\
\label{eq:Ute}
&&
+ \int \sigma_{\alpha\beta}^{(g)} \epsilon_{\alpha\beta}^{(edge)} dV
+ \frac{1}{2} \int \sigma_{\alpha\beta}^{(g)} \epsilon_{\alpha\beta}^{(g)} dV 
\end{eqnarray}
where $\sigma_{\alpha\beta}$ and $\epsilon_{\alpha\beta}$ are, respectively,
the stress and strain with superscripts $(edge)$, $(screw)$, and $(g)$
indicating the edge and screw components of the elastic field due to the
Shockley partial dislocation and that due to gravity, respectively.
In the above expansion diagonal relation between $(screw)$ and $(g)$
as well as $(edge)$ has been used, which shall be confirmed directly by
seeing the component of the elastic fields shown in the preceding section.
To note the fact that the sum of first two term in Eq.~(\ref{eq:Ute})
equals $U_{el}$ puts the present issue in relief;
only the term $\int \sigma_{\alpha\beta}^{(g)} \epsilon_{\alpha\beta}^{(edge)} dV$
depends simultaneously on $(g)$ and $(edge)$.
The self-energy term
$\frac{1}{2} \int \sigma_{\alpha\beta}^{(g)} \epsilon_{\alpha\beta}^{(g)} dV$
of the elastic field due to gravity does not depend on the position of the
Shockley partial dislocation.

\section{Calculation of cross term \label{sec:crossterm}}
\begin{figure}[t]
\centerline{
\epsfxsize=0.45\textwidth
\epsfbox{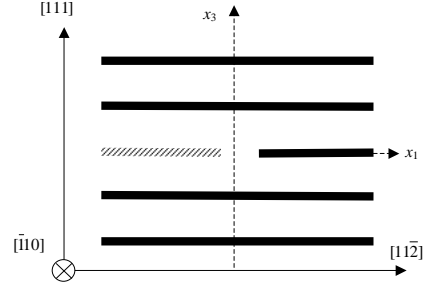}}
\caption{\label{fig:coordinate}
The coordination system $x_1x_2x_3$ for calculation of the elastic field
due to the Shockley partial dislocation.
The hexagonal planes are shown by thick horizontal lines.
The intrinsic stacking fault is hatched.
$x_3$ axis is taken to be perpendicular to the hexagonal planes ($[111]$),
$x_2$ axis along the dislocation line ($[\bar{1}10]$),
and $x_1$ axis parallel to the stacking fault ($[11\bar{2}]$).}
\end{figure}
At first, we write the elastic fields due to gravity and the Shockley partial
dislocation.
Integrating
\begin{equation}
\label{eq:fg}
f_{\alpha}^{(g)} = \frac{\partial \sigma_{\alpha\beta}^{(g)}}{\partial x_\beta},
\end{equation}
with the gravitational force $f^{(g)}$ equaling the gradient of the pressure $P$
[as shown by Eq.~(\ref{eq:mechanical}) the $x$ and $y$ components vanish]
we have
\begin{equation}
\label{eq:sigmag}
\sigma_{zz}^{(g)} = \sigma_{zz}^{(g)}(0) - mg\rho(0)\frac{1}{\xi}\ln(1+\xi z).
\end{equation}
Here, we have used the relation $\rho(z)=\rho(0)/[1+\xi z]$ [insertion of
Eq.~({\ref{eq:c}}) into $\rho(z)=4/abc(z)$].
The stress components $\sigma_{\alpha\beta}^{(g)}$ except for $\alpha\beta=zz$
vanish.
Accordingly, we have
\begin{equation}
\label{eq:integral}
\int \sigma_{\alpha\beta}^{(g)} \epsilon_{\alpha\beta}^{(edge)} dV
=
\int \sigma_{zz}^{(g)} \epsilon_{zz}^{(edge)} dV.
\end{equation}
To calculate $\epsilon_{zz}^{(edge)}$ as done previously~\cite{Mori2009}
we introduce coordination system $x_1x_2x_3$ shown in Fig.~3.
$x_3$ axis is taken to be perpendicular to the hexagonal plane,
$x_2$ axis paral{\bf lel}
to the dislocation line
($(1/\sqrt{2})[\bar{1}10]$),
and $x_1$ axis [within (111) and perpendicular to $x_2$]
so as to the coordination system to be a right-handed system.
Thus, we have unit vectors pointing to the axes as
\begin{equation}
\hat{x_1} = \frac{1}{\sqrt{6}}[11\bar{2}],
\hat{x_2} = \frac{1}{\sqrt{2}}[\bar{1}10],
\hat{x_3} = \frac{1}{\sqrt{3}}[111].
\end{equation}
In the frame of $x_1x_2x_3$ the non-vanishing component of the strain
due to the Shockley partial dislocation with Burgers vector
$\mbox{\boldmath $b$}^I = \mbox{\boldmath $b$}_{\parallel}^I
+ \mbox{\boldmath $b$}_{\perp}^I$
is given as \cite{Hirth}
\begin{eqnarray}
&&
\epsilon_{11}^{(edge)} = -\frac{|b_{\perp}^I|}{2\pi}
\frac{x_3[(2\lambda+3\mu)x_1^2+\mu x_3^2]}{(\lambda+2\mu )(x_1^2+x_3^2)^2}, \\
&&
\epsilon_{33}^{(edge)} = \frac{|b_{\perp}^I|}{2\pi}
\frac{x_3[(2\lambda+\mu)x_1^2-\mu x_3^2]}{(\lambda+2\mu )(x_1^2+x_3^2)^2}, \\
&&
\epsilon_{13}^{(edge)}=\epsilon_{31}^{(edge)} = \frac{|b_{\perp}^I|}{4\pi (1-\nu)}
\frac{x_1(x_1^2-x_3^2)}{(x_1^2+x_3^2)^2}, \hspace{1cm} \\
&&
\epsilon_{12}^{(screw)} = \epsilon_{21}^{(screw)} = -\frac{|b_{\parallel}^I|}{4\pi}
\frac{x_3}{x_1^2+x_3^2}, \\
&&
\epsilon_{23}^{(screw)} = \epsilon_{32}^{(screw)} = \frac{|b_{\parallel}^I|}{4\pi}
\frac{x_1}{x_1^2+x_3^2},
\end{eqnarray}
where $\mbox{\boldmath $b$}_{\parallel}^I$
and $\mbox{\boldmath $b$}_{\perp}^I$
are, respectively, the parallel (screw) and normal (edge) components
of $\mbox{\boldmath $b$}^I$ to the dislocation line.
The stress components are readily obtained \cite{Hirth}.
As mentioned previously we confirm the orthogonal relation, i.e.,
$\sigma_{ij}^{(edge)} \epsilon_{ij}^{(screw)}$ = 0.
Here and hereafter, subscripts $i,j$ represent $1,2,3$.
The orthogonal relation between $(g)$ and $(screw)$ is confirmed
by coordinate transformation between $xyz$ and $x_1x_2x_3$.
This coordinate transformation is also necessary to calculate
$\int \sigma_{\alpha\beta}^{(g)} \epsilon_{\alpha\beta}^{(edge)} dV$.

The matrix of the coordinate transformation from $x_1x_2x_3$ system
to $xyz$ system is given by the inner product of the axis vectors
$\hat{x_\alpha} \cdot \hat{x_i}$.
The result is as
\begin{equation}
a_{\alpha i} = \left(
\begin{array}{ccc}
a_{x1} & a_{x2} & a_{x3} \\
a_{y1} & a_{y2} & a_{y3} \\
a_{z1} & a_{z2} & a_{z3}
\end{array}
\right)
=\left(
\begin{array}{ccc}
\frac{1}{\sqrt{3}} & 0 & \frac{2}{\sqrt{6}} \\
0 & 1 & 0 \\
-\frac{2}{\sqrt{6}} & 0 & \frac{1}{\sqrt{3}}
\end{array}
\right),
\end{equation}
which is the mere rotation in $xz$ plane.
We calculate $\epsilon_{zz}^{(edge)}$ as follows.
\begin{eqnarray}
\nonumber
&&
\epsilon_{zz}^{(edge)} =
a_{zi} a_{zj} \epsilon_{ij}^{(edge)} \\
\nonumber
&&
= a_{z1} a_{z1} \epsilon_{11}^{(edge)} + 2 a_{z1} a_{z3} \epsilon_{13}^{(edge)}
+ a_{z3} a_{z3} \epsilon_{33}^{(edge)} \\
\nonumber
&&
= \frac{|\mbox{\boldmath $b$}_{\perp}^I|x_3}{6\pi(\lambda+2\mu)(x_1^2+x_3^2)^2}
[(-2\lambda-5\mu)x_1^2 -3\mu x_3] \\
&&
-\frac{|\mbox{\boldmath $b$}_{\perp}^I|}{3\sqrt{2}\pi(1-\nu)}
\frac{x_1(x_1^2-x_3^2)}{(x_1^2+x_3^2)^2}.
\end{eqnarray}
By calculation $\epsilon_{zz}^{(screw)}$ vanishes.

Now, let us transform
$\left(
\begin{array}{c}
x_1 \\ x_2 \\ x_3
\end{array}
\right)$
into
$\left(
\begin{array}{c}
x \\ y \\ z
\end{array}
\right)$.
The inverse transform of $x_\alpha = a_{\alpha i} x_i$ is given by
the transpose of $a_{\alpha i}$ because $a_{\alpha i}$ is orthogonal.
\begin{eqnarray}
\nonumber
\left(
\begin{array}{c}
x \\ y \\ z
\end{array}
\right)
=
\left(
\begin{array}{ccc}
\frac{1}{\sqrt{3}} & 0 & \frac{2}{\sqrt{6}} \\
0 & 1 & 0 \\
-\frac{2}{\sqrt{6}} & 0 & \frac{1}{3}
\end{array}
\right)
\left(
\begin{array}{c}
x_1 \\ x_2 \\ x_3
\end{array}
\right), && \\
\left(
\begin{array}{c}
x_1 \\ x_2 \\ x_3
\end{array}
\right)
=
\left(
\begin{array}{ccc}
\frac{1}{\sqrt{3}} & 0 & -\frac{2}{\sqrt{6}} \\
0 & 1 & 0 \\
\frac{2}{\sqrt{6}} & 0 & \frac{1}{3}
\end{array}
\right)
\left(
\begin{array}{c}
x \\ y \\ z
\end{array}
\right). &&
\end{eqnarray}
$\epsilon_{zz}^{(edge)}$ is rewritten in terms of $x$,$y$, and $z$ as
\begin{eqnarray}
\nonumber
&&
\epsilon_{zz}^{(edge)} =
\frac{\mbox{\boldmath $b$}_{\perp}^I}{18\sqrt{3}\pi(x^2+z^2)^2}
\left\{
\frac{1}{\lambda +2\mu}
\right. \\
\nonumber
&&
\times \left[
-\sqrt{2}(2\lambda+11\mu )x^3 + (6\lambda-3\mu )x^2z
\right. \\
\nonumber
&&
\left.
-9\sqrt{2}\mu xz^2 - (4\lambda+13\mu)z^3
\right] \\
\nonumber
&&
+\frac{1}{1-\nu} \left(
\sqrt{2}x^3 + 6x^2z
\right. \\
\label{eq:epsilon}
&&
\left.\left.
-5\sqrt{2}xz^2 + 2z^3
\right)
\right\}.
\end{eqnarray}

Let us calculate
$\int \sigma_{zz}^{(g)} \epsilon_{zz}^{(edge)} dV$
substituting by Eqs.~(\ref{eq:sigmag}) and (\ref{eq:epsilon}).
Integration will be performed in the cylindrical coordinate $(r,\phi)$
with the cylindrical axis being $y$ axis.
$r$ spans $[r_0,R]$ and $\phi$ does $[0,2\pi]$.
At first, $\int \sigma_{zz}^{(g)}(0) \epsilon_{zz}^{(edge)} dV = 0$
is shown by simple calculation.
So, we have
\begin{equation}
\label{eq:lnepsilon}
\int \sigma_{zz}^{(g)} \epsilon_{zz}^{(edge)} dV
=
- mg\rho(0)\frac{1}{\xi} \int \ln(1+\xi z) \epsilon_{zz}^{(edge)} dV.
\end{equation}
The integrations including $\ln(1+\xi z)$ are shown in
Appendix.
We perform the integration for $r$, and then expanding with respect to
$\xi$ and integrate for $\phi$.
In expansion $\xi R$ is treated as a small quantity.
In Ref.~\cite{Mori2006STAM} $\xi$ is of the order of $10^{-3}$;
so, for the grain with linear dimension $R$ of several hundreds
of particles this assumption is justified.

\begin{figure}[t]
\centerline{
\epsfxsize=0.45\textwidth
\epsfbox{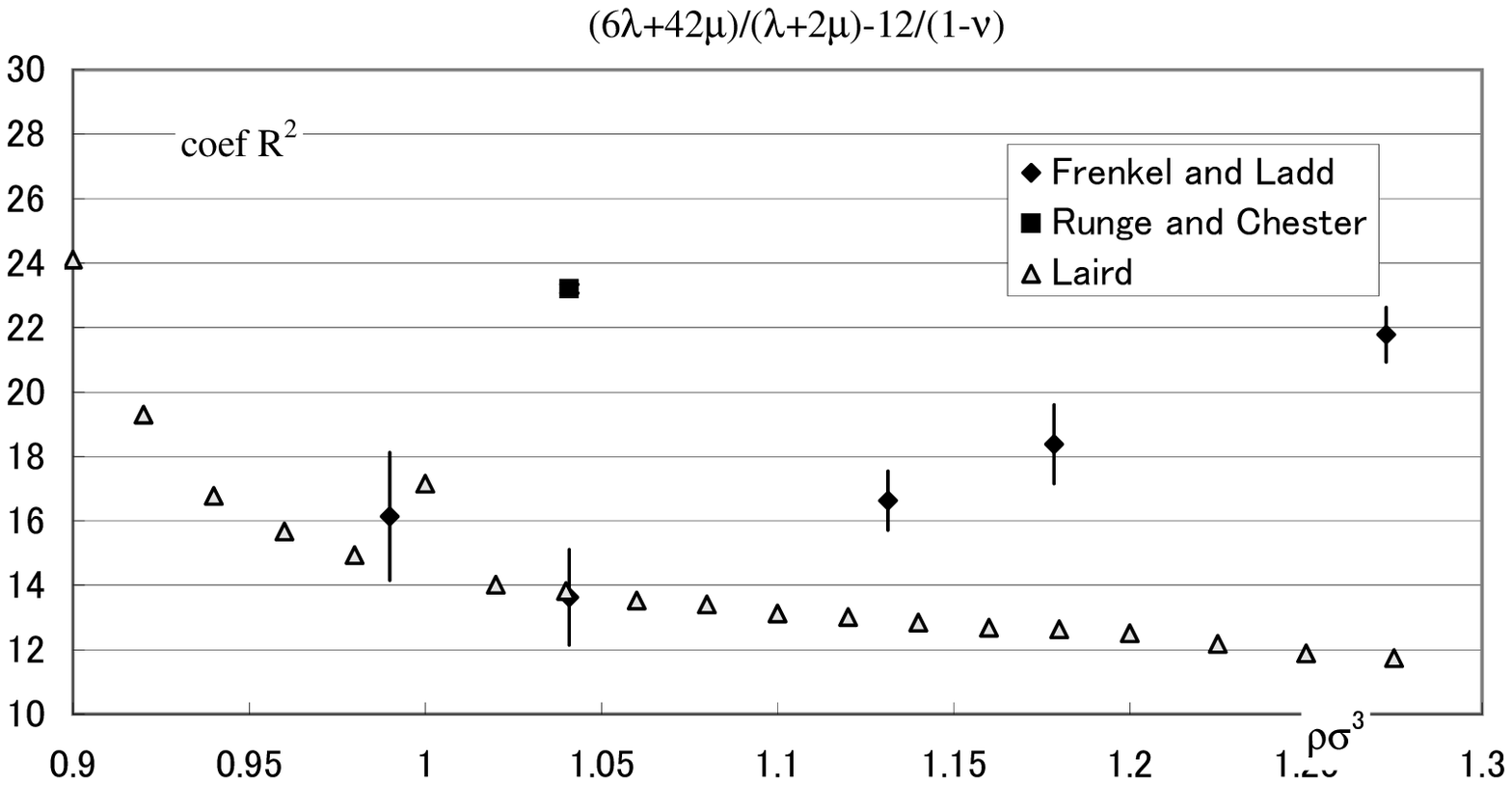}}
\centerline{
\epsfxsize=0.45\textwidth
\epsfbox{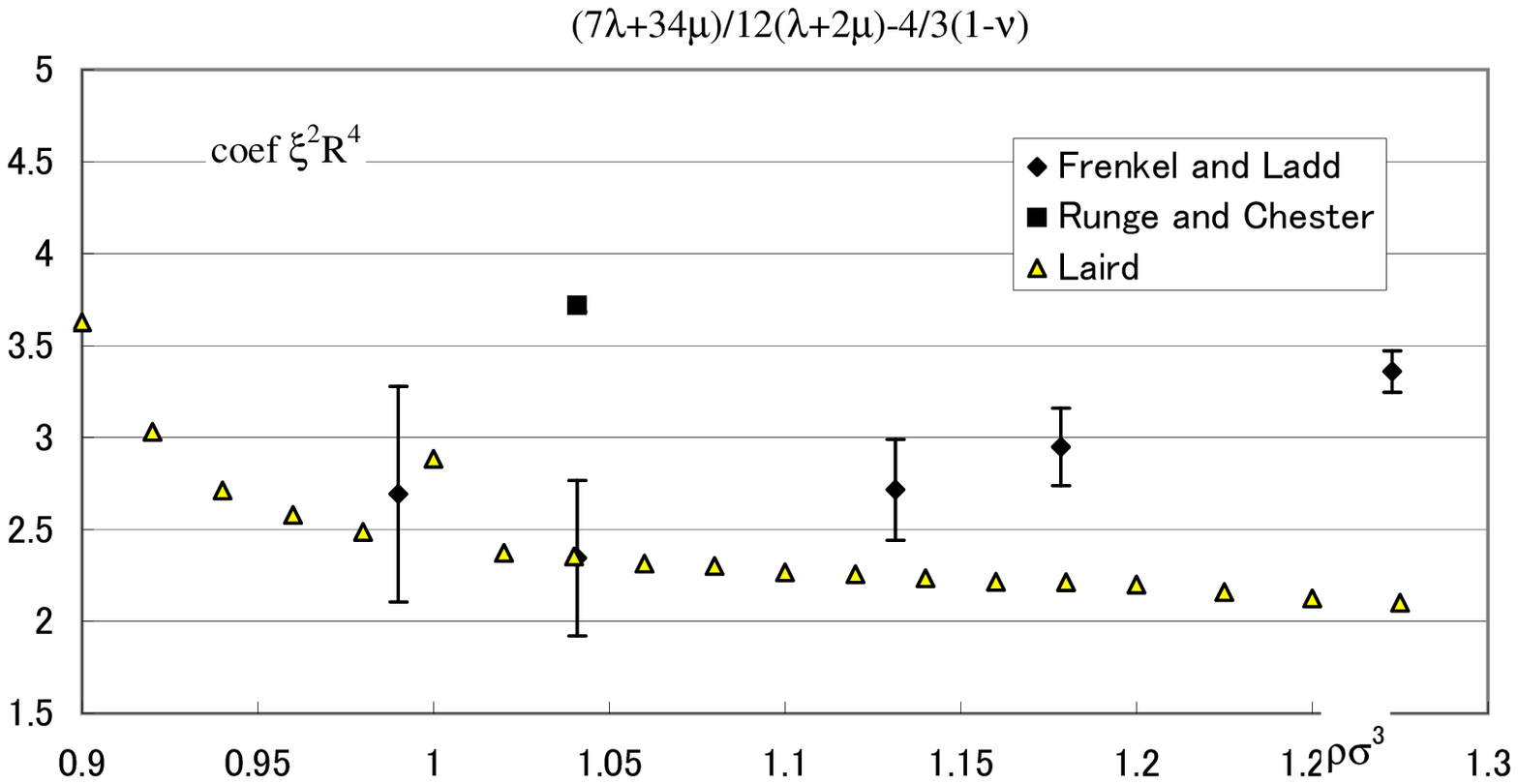}}
\caption{\label{fig:coef}
The coefficients of $R^2$ (top) and $R^4$ (bottom) of the cross-coupling
term [Eqs.~(\ref{eq:const}) and (\ref{eq:coef})] are plotted
against $\rho\sigma^3$ using values of elastic constants calculated in
Refs.~\cite{Frenkel1987,Runge1987,Laird1992}.
Both coefficients are shown to be positive.}
\end{figure}

Substituting by Eq.~(\ref{eq:epsilon}) in Eq.~(\ref{eq:lnepsilon})
and then using Eqs.~(\ref{eq:resx3ln}-\ref{eq:resz3ln}) we obtain
\begin{eqnarray}
\nonumber
&&
\int\sigma_{zz}^{(g)}\epsilon_{zz}^{(edge)} dV
=
\frac{|\mbox{\boldmath $b$}_\perp^I|mg\rho(0)}{144\sqrt{3}}
\\
\nonumber
&&
\times \left[
\frac{6\lambda+42\mu}{\lambda+2\mu}
+\frac{14\lambda+68\mu}{\lambda+2\mu} \frac{\xi^2(R^2+r_0^2)}{12}
\right.
\\
\nonumber
&&
\left.
-\frac{1}{1-\nu}
\left(
12+\frac{2\xi^2(R^2+r_0^2)}{3}
\right)
\right] \\
\label{eq:final}
&&
\times (R^2-r_0^2).
\end{eqnarray}
Let us look at the mangitude of the coefficients of $R^2$ and $R^4$.
In the square bracket in Eq.~(\ref{eq:final}) the constant term is
\begin{equation}
\label{eq:const}
\frac{6\lambda+42\mu}{\lambda+2\mu}
-\frac{12}{1-\nu}
\mbox{\bf ,}
\end{equation}
and the coefficient of $\xi^2(R^2+r_0^2)$ is
\begin{equation}
\label{eq:coef}
\frac{14\lambda+68\mu}{12(\lambda+2\mu)}
-\frac{4}{3(1-\nu)}.
\end{equation}
Quantities of Eqs.~(\ref{eq:const}) and (\ref{eq:coef}) are plotted against
$\rho\sigma^3$ in
Fig.~\ref{fig:coef}.
It is shown that both coefficients are positive, indicating that
the cross-coupling term also yields a driving force for the Shockley
partial dislocation moving toward the upper boundary of the grain. 

\section{Concluding remarks \label{sec:conc}}
We have considered the system including an intrinsic stacking fault
with a Shockley partial dislocation terminating its lower end under
gravity.
In addition to the self energy of the elastic field due to the Shockley
partial dislocation, which has been shown to yield a driving force
for the Shockley partial dislocation moving toward the upper boundary
of the grain \cite{Mori2009},
the cross-coupling term between the elastic field due to gravity
and that due to the Shockley partial dislocation has been calculated.
Driving force for the Shockley partial dislocation moving toward the
upper boundary of the grain is also shown to arise from the cross term.

The coefficient of $\ln R$ in Eq.~(\ref{eq:Uel}) is, at most, 5.
Taking into account the denominator, the coefficient of $R^2$
in Eq.~(\ref{eq:final}) is a little less than this coefficient.
It shows that the cross-coupling term gives a contribution,
which is not negligible.
In contrast, the coefficient of $R^4$ is negligibly small because of
the prefactor of $\xi^2$.
Of course, the magnitude of the coefficient $\xi$ depends on the strength
of gravity (on $m\sigma$ and $1/k_BT$ through $mg\sigma/k_BT$
in natural sedimentation, and directly on $g$ in centrifugal sedimentation
\cite{Megens1997,Ackerson1999,Suzuki2007}).
Under an extremely high $g$ condition, however, the linear relation
between $c$ and $z$, Eq.~(\ref{eq:c}), may not hold.
This regime is, nevertheless, interesting
because more enhancement of gravitational effect is expected.
Consideration on this regime is left as a future research.

We wish to emphasize that the simulation of colloidal epitaxy is underway.
The epitaxial growth has been confirmed when the lateral system size
was increased doubly and doubly.
It is an indication that the stress causing the epitaxial growth
is realizable.
Not only simulations but also elastic theoretical study on the effect
of the patterned substrate under gravity is also a future theme.

\appendix
\section{Calculations of integrals}
Substitution of Eq.~(\ref{eq:epsilon}) in the integral
in Eq.~(\ref{eq:lnepsilon}) arises the following integrals.
\begin{eqnarray}
\nonumber
&&
\int \frac{x^3\ln(1+\xi z)}{(x^2+z^2)^2} dxdz
=
\int_{r_0}^{R}\int_0^{2\pi} \cos^3\phi \\
\label{eq:x3ln}
&&
\times \ln (1+\xi r\sin\phi) drd\phi, \\
\nonumber
&&
\int \frac{x^2z\ln(1+\xi z)}{(x^2+z^2)^2} dxdz
=
\int_{r_0}^{R}\int_0^{2\pi} \cos^2\phi\sin\phi \\
&&
\times \ln (1+\xi r\sin\phi) drd\phi, \\
\nonumber
&&
\int \frac{xz^2\ln(1+\xi z)}{(x^2+z^2)^2} dxdz
=
\int_{r_0}^{R}\int_0^{2\pi} \cos\phi\sin^2\phi \\
&&
\times \ln (1+\xi r\sin\phi) drd\phi, \\
\nonumber
&&
\int \frac{z^3\ln(1+\xi z)}{(x^2+z^2)^2} dxdz
=
\int_{r_0}^{R}\int_0^{2\pi} \sin^3\phi \\
\label{eq:z3ln}
&&
\times \ln (1+\xi r\sin\phi) drd\phi.
\end{eqnarray}
Integration for $r$ can be done by applying
\begin{eqnarray}
\nonumber
\int \ln(1+\xi r \sin\phi)dr =
\frac{1}{\xi \sin\phi}(1+\xi r\sin\phi) & & \\
\times \ln(1+\xi r\sin\phi)- r. & &
\end{eqnarray}
Right-hand sides (RHS) of Eqs.~(\ref{eq:x3ln}-\ref{eq:z3ln}) are
calculated to be
\begin{eqnarray}
\nonumber
&&
\int\frac{x^3\ln(1+\xi z)}{(x^2+z^2)^2} dxdz =
\frac{1}{\xi} \int_0^{2\pi} \frac{\cos^3\phi}{\sin\phi} \\
\nonumber
&&
\times \left[
(1+\xi R\sin\phi)\ln(1+\xi R\sin\phi)
\right. \\
&&
\left.
-(1+\xi r_0\sin\phi)\ln(1+\xi r_0\sin\phi)
\right] d\phi, \\
\nonumber
&&
\int\frac{x^2z\ln(1+\xi z)}{(x^2+z^2)^2} dxdz =
\frac{1}{\xi} \int_0^{2\pi} \cos^2\phi \\
\nonumber
&&
\times \left[
(1+\xi R\sin\phi)\ln(1+\xi R\sin\phi)
\right. \\
&&
\left.
-(1+\xi r_0\sin\phi)\ln(1+\xi r_0\sin\phi)
\right] d\phi, \\
\nonumber
&&
\int\frac{xz^2\ln(1+\xi z)}{(x^2+z^2)^2} dxdz =
\frac{1}{\xi} \int_0^{2\pi} \cos\phi\sin\phi \\
\nonumber
&&
\times \left[
(1+\xi R\sin\phi)\ln(1+\xi R\sin\phi)
\right. \\
&&
\left.
-(1+\xi r_0\sin\phi)\ln(1+\xi r_0\sin\phi)
\right] d\phi, \\
\nonumber
&&
\int\frac{z^3\ln(1+\xi z)}{(x^2+z^2)^2} dxdz =
\frac{1}{\xi} \int_0^{2\pi} \sin^2\phi \\
\nonumber
&&
\times \left[
(1+\xi R\sin\phi)\ln(1+\xi R\sin\phi)
\right. \\
&&
\left.
-(1+\xi r_0\sin\phi)\ln(1+\xi r_0\sin\phi)
\right] d\phi.
\end{eqnarray}
Here, $\int_0^{2\pi} \cos^3\phi d\phi =0$,
$\int_0^{2\pi} \cos^2\phi\sin\phi d\phi =0$,
$\int_0^{2\pi} \cos\phi\sin^2\phi d\phi =0$, and
$\int_0^{2\pi} \sin^3\phi d\phi =0$
have already been used.
Let us expand $(1+a\sin\phi)\ln(1+a\sin\phi)$ with
$a$ representing $\xi R$ and $\xi r_0$.
\begin{eqnarray}
\nonumber
&&
(1+a\sin\phi)\ln(1+a\sin\phi)
= a\sin\phi + \frac{a^2}{2}\sin^2\phi \\
\label{eq:expansion}
&&
-\frac{a^3}{6}\sin^3\phi
+\frac{a^4}{12} \sin^4\phi
\cdots.
\end{eqnarray}
(The exact form is
$(1+X)\ln(1+X)-X = \sum_{n=1}^\infty \frac{(-X)^{n+1}}{n(n+1)}$.)

Integrals of $\cos^3\phi/\sin\phi$, $\cos^2\phi$,
$\cos\phi\sin\phi$, $\sin^2\phi$ multiplied by $\sin\phi$ in the
first term in RHS of Eq.~(\ref{eq:expansion}) are shown to vanish
by simple calculations.
Also, integrals of these terms multiplied by $\sin^3\phi$ in the third
term are turned to vanish.
And, for $\cos^3\phi/\sin\phi$ and $\cos\phi\sin\phi$
integrals after multiplying by $\sin^2\phi$ in the second term
and $\sin^4\phi$ in the fourth term also vanish.
Thus, we have
\begin{eqnarray}
\nonumber
&&
\int_0^{2\pi} \frac{\cos^3\phi}{\sin\phi}
(1+a\sin\phi)\ln(1+a\sin\phi) d\phi \\
\nonumber
&&
= a\int_0^{2\pi} \cos^3\phi d\phi
+ \frac{a^2}{2} \int_0^{2\pi} \cos^3\phi\sin\phi d\phi \\
\nonumber
&&
- \frac{a^3}{6} \int_0^{2\pi} \cos^3\phi\sin^2\phi d\phi
+ \frac{a^4}{12} \int_0^{2\pi} \cos^3\phi\sin^3\phi d\phi \\
&& = 0, \\
\nonumber
&&
\int_0^{2\pi} \cos^2\phi
(1+a\sin\phi)\ln(1+a\sin\phi) d\phi \\
\nonumber
&&
= a\int_0^{2\pi} \cos^2\phi\sin\phi d\phi
+ \frac{a^2}{2} \int_0^{2\pi} \cos^2\phi\sin^2\phi d\phi \\
\nonumber
&&
- \frac{a^3}{6} \int_0^{2\pi} \cos^2\phi\sin^3\phi d\phi
+ \frac{a^4}{12} \int_0^{2\pi} \cos^2\phi\sin^4\phi d\phi \\
&& = 0 + \frac{a^2}{2}\frac{\pi}{4} + 0 + \frac{a^4}{12}\frac{\pi}{8}
\cdots, \\
\nonumber
&&
\int_0^{2\pi} \cos\phi\sin\phi
(1+a\sin\phi)\ln(1+a\sin\phi) d\phi \\
\nonumber
&&
= a\int_0^{2\pi} \cos\phi\sin^2\phi d\phi
+ \frac{a^2}{2} \int_0^{2\pi} \cos\phi\sin^3\phi d\phi \\
\nonumber
&&
- \frac{a^3}{6} \int_0^{2\pi} \cos\phi\sin^4\phi d\phi
+ \frac{a^4}{12} \int_0^{2\pi} \cos\phi\sin^5\phi d\phi \\
&& = 0, \\
\nonumber
&&
\int_0^{2\pi} \sin^2\phi
(1+a\sin\phi)\ln(1+a\sin\phi) d\phi \\
\nonumber
&&
= a\int_0^{2\pi} \sin^3\phi d\phi
+ \frac{a^2}{2} \int_0^{2\pi} \sin^4\phi d\phi \\
\nonumber
&&
- \frac{a^3}{6} \int_0^{2\pi} \sin^5\phi d\phi
+ \frac{a^4}{12} \int_0^{2\pi} \sin^4\phi d\phi \\
&& = 0 + \frac{a^2}{2}\frac{3\pi}{4} + 0 + \frac{a^4}{12}\frac{5\pi}{8}
\cdots.
\end{eqnarray}
Accordingly, up to the third order in $\xi$ the following results are obtained.
\begin{eqnarray}
\label{eq:resx3ln}
&&
\int \frac{x^3\ln(1+\xi z)}{(x^2+z^2)^2} dxdz = 0, \\
\nonumber
&&
\int \frac{x^2z\ln(1+\xi z)}{(x^2+z^2)^2} dxdz =
\frac{\xi(R^2-r_0^2)\pi}{8} \\
&&
+ \frac{\xi^3(R^4-r_0^4)\pi}{96}, \\
&&
\int \frac{xz^2\ln(1+\xi z)}{(x^2+z^2)^2} dxdz = 0, \\
\nonumber
&&
\int \frac{z^3\ln(1+\xi z)}{(x^2+z^2)^2} dxdz =
\frac{3\xi(R^2-r_0^2)\pi}{8} \\
\label{eq:resz3ln}
&&
+ \frac{5\xi^3(R^4-r_0^4)\pi}{96}.
\end{eqnarray}

\end{document}